\documentclass[aps,twocolumn,showpacs,superscriptaddress,groupedaddress]{revtex4-1}
\usepackage[latin1]{inputenc}
\usepackage{amsmath}
\usepackage{amssymb}
\usepackage{graphicx}  
\usepackage{dcolumn}   
\usepackage{bm}        
\usepackage{array}
\usepackage{color}
\usepackage{hyperref}
\hypersetup{colorlinks=true,%
linkcolor=blue,%
pdfproducer={Birds-of-a-Feather},%
pdfstartview=FitB}
\newlength{\figwidth}
\usepackage{epsf}

\DeclareMathOperator{\Tr}{Tr}
\definecolor{darkgreen}{rgb}{0.,0.51,0.}

\begin{document}
\title{Time delay matrix at the spectrum edge and the minimal chaotic cavities}

\author{Adel Abbout}

\affiliation{Service de Physique de l'\'Etat Condens\'e (CNRS URA 2464),
IRAMIS/SPEC, CEA Saclay, 91191 Gif-sur-Yvette, France.\\
Laboratoire CRISMAT, UMR 6508 CNRS, ENSICAEN et Universit\'e de Caen Basse Normandie, 6 Boulevard Mar\'echal Juin,
F-14050 Caen, France.}
   
\begin{abstract}

Using the concept of minimal chaotic cavities, we give the distribution of the proper delay times of $Q=-i\hbar \mathcal{S}^\dagger\frac{\partial \mathcal{S}}{\partial E}$ at the spectrum edge with a scattering matrix  $\mathcal{S}$ belonging to  circular ensembles CE. The three classes of symmetry ($\beta=1,2 $ and $ 4$)
 will be analyzed to  show  how  it differs from the distribution obtained in the bulk of the spectrum. In this new class of universality at the spectrum edge, more attention will be given to the  Wigner's time
$\tau_w=\Tr(Q)$ and its distribution will be given analytically in the case of 2 modes scattering. The results will be presented exactly at all the Fermi energies without any approximation. All this will be tested numerically with an excellent precision.

\end{abstract}
\maketitle

Since it was introduced by Eisenbud and Wigner\cite{Eisenbud}, the time delay in scattering  problems was extensively studied and used in describing  quantum transport. This quantity was generalized by Smith to the notion
 of time delay matrix $Q\equiv -i\hbar \mathcal{S}^\dagger\frac{\partial \mathcal{S}}{\partial E}$ in multichannel scattering\cite{Smith}. The eigenvalues of the matrix $Q$, known as the
 \textit{proper delay times}  play an important role in considering interaction in low dimensional systems such as mesoscopic capacitors \cite{MelloButtiker} and quantum dots \cite{BrouwerVanLangen}. The delay times took
more attention in chaotic systems and random potentials to provide a statistical description of the time spent by wavepakets in the interacting region and the fluctuation of physical observables directly related to, such as
thermopower\cite{1stArticle}. The distribution of the delay time  was obtained in the case of  one-dimensional disordered systems
by means of  invariant embedding formalism in \cite{Kumar} and also in \cite{Texier}. The more general result   including the different classes of symmetry is obtained in \cite{MelloButtiker} but still for 1D systems (one mode). The next step
of treating the N-modes scattering problems, is more complicated. However, the authors of \cite {Fyodorov} succeeded and gave the distribution
of individual \textit{partial} delay times defined as the energy derivative of the phase shifts $\tau_i^{partial}\equiv\frac{ \partial \theta_i}{\partial E}$. The joint probability distribution of the proper delay times was finally obtained in \cite{BrouwerFrahm} which, expressed in terms of the inverse proper delay times $\gamma_i=\frac{1}{\tau_i}$, reads:

\begin{eqnarray}
\mathcal{P}(\{\gamma_n\})\propto \prod_{i<j} |\gamma_i-\gamma_j|^\beta \prod_k \gamma_k^{\beta M/2} e^{-\pi \beta \gamma_k/\Delta} \label{TimeDelay}
\end{eqnarray}
$\beta$ stands for the symmetry class($\beta=1$, under time reversal symmetry (TRS), $\beta=2$ broken TRS, and $\beta=4$ broken spin-rotation symmetry ). $\Delta$ is the mean level spacing.
 We need to mention that this result is obtained for large Hamiltonians($N\rightarrow\infty$) and  at Fermi energy lying in the bulk spectrum($\Delta\sim 1/N$)\cite{Hackenbroich}.
 In this paper, we give the distribution of $Q$ at the edge of the spectrum, for small values of the Hamiltonian size. To do this, we start with the following representation of the scattering matrix:
\begin{eqnarray}
\mathcal{S}=[I-i\pi \mathcal{W}^\dagger R \mathcal{W} ]\times [I+i\pi \mathcal{W}^\dagger R \mathcal{W} ]^{-1} \label{Parametrisation}
\end{eqnarray}
\begin{figure}
\begin{center}
\includegraphics[scale=0.17]{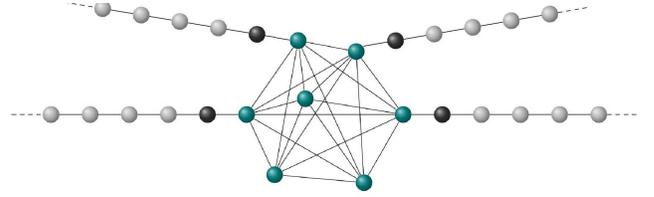}
\end{center}
\caption{Tight binding model for a chaotic cavity with $N=7$ sites and $M=4$ leads. In order to obtain Scattering matrix from circular ensembles CE, $N$ should be larger or equal to $M$. The case 
$N=M$ is referred here as a minimal chaotic cavity. }
\label{Cavity}

\end{figure}
where we introduced the reaction matrix $R=(E-\Lambda-\mathcal{H})^{-1}$. $\mathcal{H}$ is an $N\times N$ Hamiltonian  matrix describing the closed chaotic system
and $\mathcal{W}$ is $M\times N$ coupling matrix between the cavity and the conducting modes. This way, the matrix $\Sigma=\Lambda-i\pi \mathcal{W}\mathcal{W}^\dagger$ represents the self-energy
of the leads (conducting modes) where the real part $\Lambda$ and the energy dependence of $\Sigma$ will not be neglected as in the wide band limit WBL which is by the way the case of Eq. (\ref{TimeDelay}) obtained
 by \cite{BrouwerFrahm}. Hereafter, the case of uniform semi-infinite leads is assumed, and this makes, with the independence  and equivalence assumption between modes,   $\Sigma=(E/2-i\sqrt{1-(E/2)^2}) \times I_{M}$\cite{Sasada}. We work also within the equal a priori probability ansatz by requiring
uniform distribution for $\mathcal{S}$ which is the definition of circular ensembles CE where $\langle \mathcal{S} \rangle=0$. Cavities with this condition will be called \textit{completely chaotic}. It is easy to show\cite{BrouwerLorentz} that if the Hamiltonian is Lorentzian with center $E-\Lambda$ and width $\Gamma/2=-\Im\Sigma$ we get $\mathcal{S}$ circular with the corresponding class of symmetry ($\beta=1,2$ and $4$)\cite{1stArticle}. It is important to mention that this is true for all $N \ge M$ so that the size of $\mathcal{H}$ is not required to be infinite and thus, all the cavities with different sizes are equivalent and lead to the same
distribution of any observable expressed using the elements of $\mathcal{S}$ such as the conductance\cite{Jalabert} or shot noise. Unfortunately, this is not true for observables containing energy derivative of $\mathcal{S}$\cite{1stArticle} so that more attention is needed. The case $N=M$ still ensures $\mathcal{S}$ from CE and we will call this case as minimal chaotic cavity. This case is very interesting since all the matrices become square and $\mathcal{S}$ is diagonalized with the same energy independent rotation that diagonalizes $\mathcal{H}$. Moreover, it allows to express $\frac{\partial\mathcal{S}}{\partial E}$ as a function of $\mathcal{S}$.
 All in all, we obtain a simple expression for $Q$ ($N=M$):
\begin{eqnarray}
   Q=\frac{\Gamma^\prime}{\Gamma} Q^- -\frac{1}{\Gamma} Q^++\frac{1}{\Gamma}, \label{MatrixDerivative}
\end{eqnarray}
where we put $Q^+=\frac{\mathcal{S}+\mathcal{S}^\dagger}{2}$ and $Q^-=\frac{\mathcal{S}-\mathcal{S}^\dagger}{2i}$. $\Gamma^\prime$ is the energy derivative of $\Gamma$. Before we give the distribution of $Q$, we need to start with the joint probability distribution j.p.d of $Q^+$ eigenvalues $\{\tau_i\}$. We use a parametrization of type Eq. (\ref{Parametrisation}) and define:
\begin{eqnarray}
P(\{\tau_i\})\propto \int  \prod_{i<j}|\epsilon_i-\epsilon_j|^\beta \prod_k \frac{\delta(\tau_k-\frac{1-\epsilon_k^2}{1+\epsilon_k^2})}{(1+\epsilon_k^2)^d}d\epsilon_k \label{Eigenenergies}
\end{eqnarray}
and $d=\beta(N-1)/2+1$. Here, the $\{\epsilon_i\}$ are the eigenvalues of the matrix $\pi \mathcal{W}^\dagger R \mathcal{W} $ which is also distributed according to Lorentzian ensembles\cite{1stArticle}\cite{BrouwerLorentz}. After straightforward  calculations based on the properties
 of the Dirac delta function, we obtain:
\begin{eqnarray}
P(\{\tau_i\})\propto \sum_{\{\dots s_k=\pm\}} \frac{\prod_{i<j}|s_i f(\tau_i,\tau_j)-s_j f(\tau_j,\tau_i)|^\beta}{\prod_i f(\tau_i,\tau_i)} \label{Qplus}
\end{eqnarray}
\begin{figure*}
\begin{center}$
\begin{array}{ccc}
\includegraphics[width=2.in]{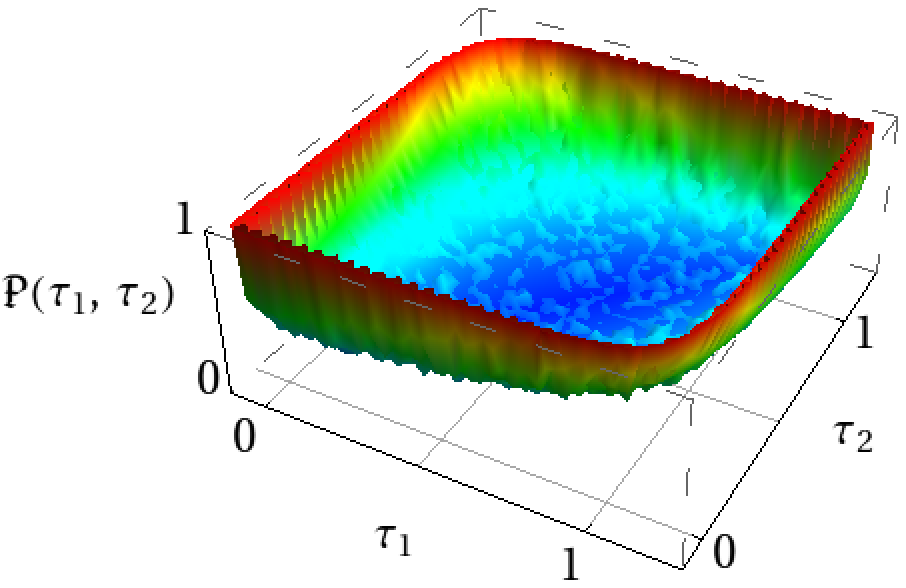} &
\includegraphics[width=2.in]{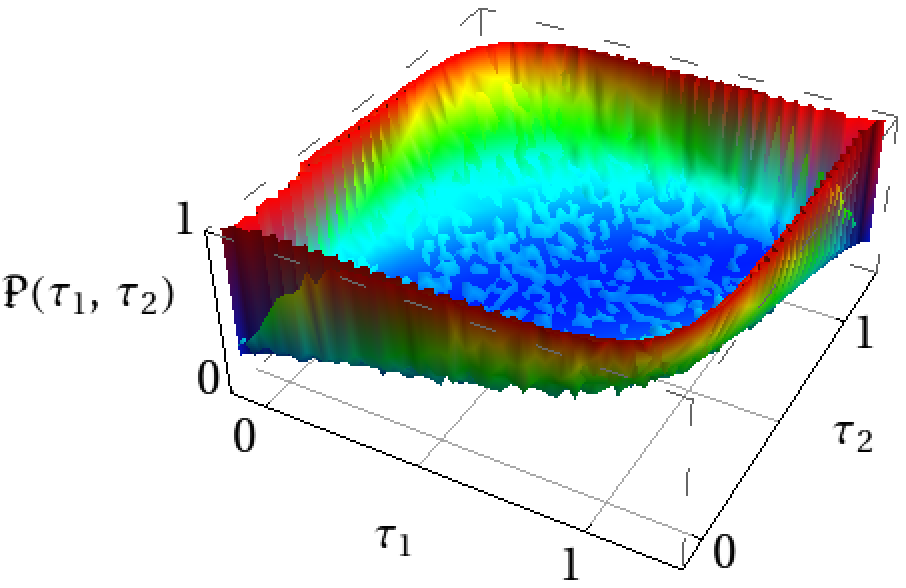} &
\includegraphics[width=2.in]{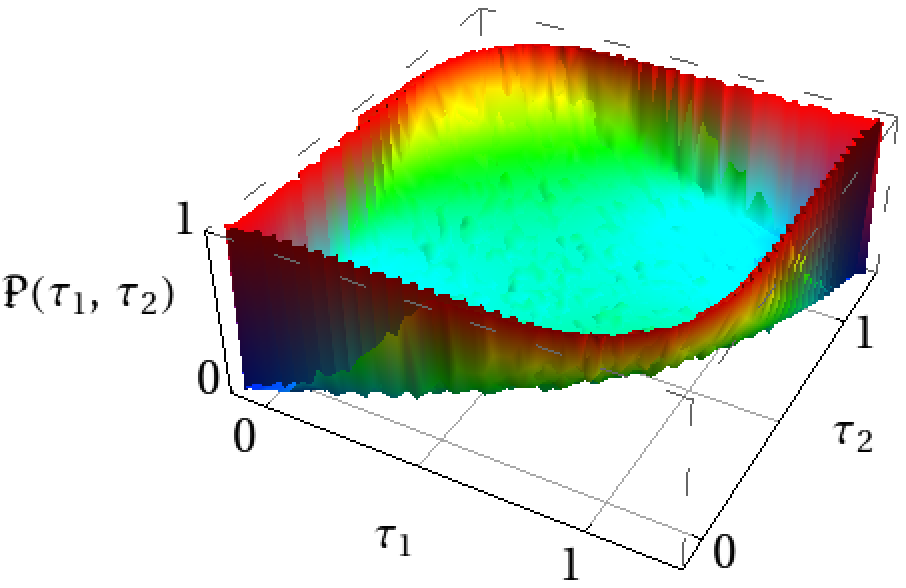}
\end{array}$
\end{center}
\caption{Joint probability distribution of the proper delay times for a minimal chaotic cavity($M=2$, $N=2$) for different symmetries: left $\beta=1$, middle $\beta=2$ and right $\beta=4$.
All the scattering matrices belong to CE. The numerical data are obtained by sampling Hamiltonians from Lorentzian ensembles. The analytical result is plotted as well 
and the matching is perfect. Here the Fermi energy $E_F=-1 t_0$.  }
\label{Threecases}
\end{figure*}
where $f(x,y)=\sqrt{(1-x)(1+y)}$ and the sum is taken over all the ensembles of $s_k$ taking values $\pm 1$. Proceeding the same way, we prove that $Q^-$ exhibits the same distribution. We remind that since $Q$, $Q^+$, $Q^-$, $S$ and $\mathcal{H}$ are diagonalized by the same rotation, the distribution of the eigenvectors is the same
as that of the eigenvectors of $\mathcal{H}$, ie uniform. That is why we restrict the study to the j.p.d of the eigenvalues. Going back to $Q$, we notice that at the half filling limit $E=0$ (the conduction band with uniform leads is $[-2t_0,2t_0]$\cite{Remark}), we have $Q=\frac{1-Q^+}{2}$ and therefore, we immediately obtain from Eq.($\ref{Qplus}$)
the j.p.d of the eigenvalues $\{\tilde{\tau}_i\}$ of $Q$ at $E=0$ just by changing in Eq. (\ref{Qplus}) $f$ by $\tilde{f}$ defined as $\tilde{f}(x,y)=\sqrt{(1-x)y}$. We need now to generalize this result to any arbitrary energy. For this task, we rewrite the matrix $Q$ at arbitrary energy as follows:
\begin{eqnarray}
Q=-\frac{\alpha}{2} \left( \frac{e^{+i\theta} \mathcal{S}+e^{-i\theta}\mathcal{S}^\dagger}{2}\right)+ \frac{1}{\Gamma} \label{GeneralQ}
\end{eqnarray}
$\alpha=(\frac{2}{\Gamma})^2$ and $\cos\theta=1/\sqrt{1+\Gamma^{\prime 2}}$, $\Gamma^\prime$ being the energy derivative of $\Gamma$.
The expression of the distribution of the matrix $Q$  comes straightforwardly and can be put in the following form:
\begin{eqnarray}
\mathcal{P}_E(Q)\propto\frac{1}{\alpha^n}\int \delta(\frac{Q-1/\Gamma+\alpha/2}{\alpha}-\left( \frac{1-\frac{\tilde{\mathcal{S}}+\tilde{\mathcal{S}}^\dagger}{2}}{2}\right))d_H \tilde{\mathcal{S}} \label{Invariance} \nonumber \\
\end{eqnarray}
where we defined here $\tilde{\mathcal{S}}= e^{+i\theta}\mathcal{S}  $. We used also in writing Eq.(\ref{Invariance}) the invariance properties  of the Haar measure
under the  transformation $\mathcal{S}\rightarrow V \mathcal{S} V^\prime$ where the arbitrary unitary matrix $V^\prime$ satisfies $V^\prime=V^T$ in presence of time
reversal symmetry $\beta=1$ and $V^\prime=V^R$ for broken spin-rotation symmetry $\beta=4$. The letters $T$, and $R$ denote respectively the transpose and the dual
operations. One can notice in Eq.(\ref{Invariance}) the presence of the expression of $Q$ at the half filling limit ($Q=\frac{1-Q^+}{2}$) this allows us to express the distribution
$\mathcal{P}_E(Q)$ at arbitrary energy as a function of the $\mathcal{P}_{E=0}(Q)$. One obtains from Eq.(\ref{Invariance}) the following formula:
\begin{eqnarray}
\mathcal{P}_E(Q)\propto\frac{1}{\alpha^N}\mathcal{P}_{E=0}\left(\frac{Q-1/\Gamma+\alpha/2}{\alpha}\right) \label{InvarianceFormula}
\end{eqnarray}
This means that all what happens at arbitrary energy in the conduction band can be obtained from the formula obtained at the half filling limit. The j.p.d of the eigenvalues $\{\tilde{\tau}_i\}$ at arbitrary energy reads:
\begin{eqnarray}
P(\{\tau_i\})\propto \sum_{\{\dots s_k=\pm\}} \frac{\prod_{i<j}|s_i f_{a,b}(\tau_i,\tau_j)-s_j f_{a,b}(\tau_j,\tau_i)|^\beta}{\prod_i f_{a,b}(\tau_i,\tau_i)}\nonumber \label{Equation9}\\
\end{eqnarray}
The function $f_{a,b}$ is defined as: $f_{a,b}(x,y)=\sqrt{(b-x)(y-a)}$. As usually for an open system, the parameters $a$ and $b$ are directly related to the conducting leads by the broadening
$\Gamma$: $a=\frac{1}{\Gamma}-\frac{\alpha}{2}$ and $b=\frac{1}{\Gamma}+\frac{\alpha}{2}$.
This result worths to stop on it to make some comments. First, we find that the repulsion term $|\tau_i-\tau_j|$ is absent in a sense that when  $\tau_i=\tau_j$ the j.p.d of the proper delay times does not vanish as observed in Eq.($\ref{TimeDelay}$). 
This can be understood if one notices that for this case ($N=M$) the transformation between the eigenenergies $\{E_i\}$ and the proper delay times $\{\tau_i\}$ is not one-to-one. So, even though  it is true that having two energies $E_i=E_j$ is forbidden by the symmetries corresponding to $\beta=1, \beta=2$ and $\beta=4$, the situation $E_i=-E_j$ is allowed, and this leads to $\tau_i=\tau_j$ .  
Second, we notice that the delay times are bounded and lie in the interval of length $\alpha$ which becomes larger and close to infinity at the bottom of the conduction band (the continuum limit). 
We can notice also that the proper delay times, as defined here, can be negative and the only situation where they are always positive is at the half filling limit where the WBL is equivalent. This causes no problem since we know that a potential can induce a delay or an advance of the wavepacket\cite{Smith}.  \\
The numerical simulations of the case of two modes are shown in Fig.\ref{Threecases}. The procedure to obtain these figures consists in sampling Hamiltonians from Lorentzian ensembles, expressing the scattering matrices using Eq. (\ref{Parametrisation}) and then calculating the corresponding 
Wigner-Smith matrix $Q$. A histogram of its eigenvalues is done at the end. The analytical result $ P(\tau_1,\tau_2)$ of Eq.(\ref{Equation9}) is plotted at the same time to check if the plotted surface passes from all the points obtained by sampling. The matching between numerical and analytical result is excellent.\\
The Hamiltonian $\mathcal{H} $ describing the chaotic cavity with a Lorentzian distribution exhibits a density of states given by $\rho(E)=\frac{N}{\pi}\frac{\Gamma/2}{(\Gamma/2)^2+(E-E_F+\Lambda)^2}$ \cite{BrouwerLorentz}\cite{AdelLorentz}. This formula simplifies too much
at the Fermi energy, where the electronic quantum transport is pertinent, to become $\rho(E_F)=\frac{\Gamma N}{2\pi}$ which means that the mean level spacing at the Fermi energy is $\Delta=\frac{2 \pi}{N \Gamma}$. 
Eq.(\ref{TimeDelay}) was obtained for a situation corresponding to $\Gamma/2=1$ and $N\rightarrow\infty$, that is to say, for a very small mean level spacing corresponding to the universality of the bulk spectrum\cite{Hackenbroich}.
In this universality class, the use of the Lorentzian ensembles is not pertinent and it can be replaced by the equivalent Gaussian ensembles\cite{BrouwerLorentz}  offering other properties such as the independence of the Hamiltonian elements.    
If we consider now a mesoscopic system with a \textit{finite number of degrees of freedom} ($N$ finite)\cite{mesoscopic}, and work with a Fermi energy lying in the bottom of the conduction band ($E_F\sim -2t_0$), the density of eigenvalues $\rho(E_F)$
tends to vanish whereas the mean level spacing $\Delta$ diverges(because $\Gamma \rightarrow0$). This situation corresponds to the edge of the spectrum with a physics different from that of the bulk. We want to show that the
system we considered in this paper ($N=M$) and we called it minimal cavity is  very important to describe this physics.\\
Suppose now that the Hamiltonian $\mathcal{H}$ of the chaotic cavity is of size $N\times N$ and that the number of conducting modes is $M<N$. We can not apply the method and the results obtained so far because 
$\mathcal{H}$ and $\mathcal{S}$ are not diagonalizable by the same set of eigenvectors anymore and moreover the matrix diagonalizing $\mathcal{S}$ is energy dependent and thus we do not obtain Eq.(\ref{MatrixDerivative}). To study such systems 
\begin{figure}
\begin{center}
\includegraphics[scale=0.45]{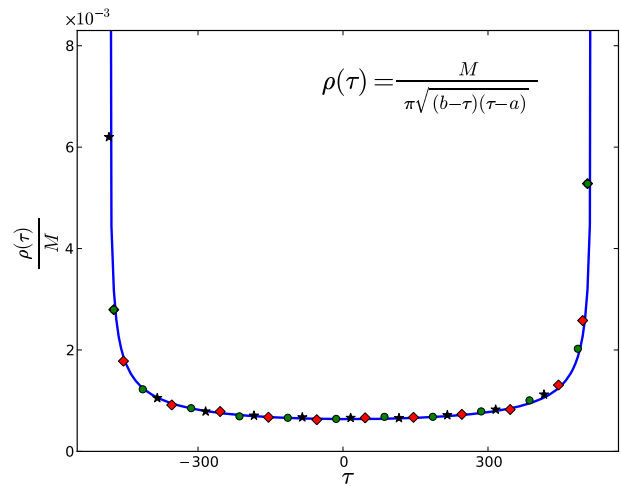}
\caption{Density of eigentimes of  chaotic cavities  with $N$ sites and $M$ leads at the spectrum edge. The scattering matrices belong to circular ensembles CE with the different 
symmetries ($\beta=1$, $\beta=2$ and $\beta=4$). {\Large $\star$} corresponds to presence of time reversal symmetry (TR) ie $\beta=1$  with $M=2$, $N=23$). {\color{red}$\blacklozenge$} is obtained in absence of (TRS) ie $\beta=2$ (with $M=9$, $N=15$).
{\large {\color{darkgreen}$\bullet$}} concerns broken spin rotational symmetry ie $\beta=4$ with ($M=5$, $N=10$). The results are obtained by sampling $\mathcal{H}$ from Lorentzian ensembles with the corresponding $\beta$-symmetry.
Fermi energy $E_F=-1.999t_0$.}
\label{density}
\end{center}
\end{figure}
we prefer to use a decimation method which consists in taking out the whole sites describing the chaotic cavity and renormalize the sites (and links between them) to which they were attached (black sites in Fig.(\ref{Cavity}))\cite{Grosso}. This is translated by taking 
a new effective  Hamiltonian $\tilde{\mathcal{H}}(E)=\hat{\tau} \frac{1}{E-\mathcal{H}} \hat{\tau}^\dagger $ with a $M\times N$ matrix $\hat{\tau}_{ij}=\delta_{ij}, i\le M $. This procedure helps in lowering the dimension
of the matrices we work with to $M\times M$ size and thus the self energy $\Sigma$ becomes proportional to identity. This procedure is exact (up to non pertinent phase factor) and ensures the same Lorentzian
 distribution for $\tilde{\mathcal{H}}$ with the same center $\mathcal{E}=E_F-\Lambda$ and width $\Gamma/2$ \cite{BrouwerLorentz}\cite{1stArticle} which means that the final $\mathcal{S}$  still belongs to CE. Replacing $\mathcal{H}$ by 
$\tilde{\mathcal{H}}(E)$ in Eq.(\ref{Parametrisation}) still do not allow to recover Eq.(\ref{MatrixDerivative}) because the new effective Hamiltonian $\tilde{\mathcal{H}}(E)$ is energy dependent. What can someone notice is that 
the derivative $\frac{\partial \tilde{\mathcal{H}}(E)}{ \partial E}=-\hat{\tau} \left(\frac{1}{E-\mathcal{H}}\right)^2 \hat{\tau}^\dagger$ becomes much more simpler at the bottom of the conduction band: In fact, we get $\Gamma \rightarrow 0$
which means a very narrow distribution, almost deterministic, and this allows to replace $\mathcal{H}$ by its mean $\mathcal{E}$  in the derivative of $\tilde{\mathcal{H}}(E)$. Moreover, $1/(E-\mathcal{E})=1/\Lambda\rightarrow -1_{M \times M}$ in 
the vicinity of the bottom of the conduction band. This result entails the following important conclusion:  
 whatever is the \textit{finite} size $N \times N$ of $\mathcal{H}$, we end with $\mathcal{S}$ and $\frac{\partial\mathcal{S}}{\partial E}$ independent of $N$ and therefore, the result obtained with the minimal 
chaotic cavities ($N=M$) applies for all of the other situations($N>M$). Moreover , this goes beyond the time delay matrix, to all the observables which can be expressed using the elements of $\mathcal{S}$ or $\frac{\partial \mathcal{S}}{\partial E}$ \cite{1stArticle}. 
We stress that, the larger is $N$ the more $\Gamma$ needs to get closer to zero. To illustrate this theory, we can look at the mean density of proper delay times for the model with $N=M$, and 
compare it to the case $N>M$ at the edge of the spectrum. This mean  density is defined as $\rho(\tau)= \sum_i \langle \delta(\tau-\tau_i)\rangle$. After straightforward calculations
we obtain the following result (see Appendix \ref{AppenA}):
\begin{eqnarray}
 \rho(\tau)=\frac{M}{\pi \sqrt{(b-\tau)(\tau-a)}}
\end{eqnarray}
 The result is $\beta$-independent which goes back to the fact that the density of eigenvalues of the Lorentzian Hamiltonian is $\beta$-independent too\cite{BrouwerLorentz}. The numerical simulations plotted in Fig.(\ref{density}) 
confirm this result and show that at the edge of the spectrum, the formula also describes the case $N>M$ as it was expected.

\textbf{The Wigner time and the  two modes scattering problem}:
The Wigner time $\tau_w$ is defined as the trace of the Wigner-Smith time delay matrix: $\tau_w\equiv\Tr(Q)$. This observable is very important since it can be directly related to 
the change in the local density of states \cite{Statofumi}\cite{Akkermans} and thus enters in the evaluation of the thermodynamic expectation of physical quantities where we take into account the effect of the coulomb 
interactions in the grand canonical average\cite{BrouwerVanLangen}. For the minimal chaotic cavity $(N=M)$, the use of the n-level correlation functions  of circular ensembles\cite{Dyson}\cite{Haake} gives access to the first cumulant 
which is found $\beta$-independent, $\langle\tau_w\rangle=\frac{M}{\Gamma}$(see appendix \ref{AppenB}) as well as the second cumulant  $\langle \delta\tau_w^2 \rangle=\frac{\alpha^2}{8} \langle|t|^2 \rangle$(see Appendix \ref{AppenC}) where $t=\Tr(S)$. For 
$\beta=1$ we get$\langle|t|^2 \rangle=\frac{2N}{N+1}$ and for $\beta=2$ we have $\langle|t|^2 \rangle=1$ \cite{Haake}. The whole distribution is 
more complex to obtain and can be expressed only in the simple cases $N=1$ and $N=2$. For $N=1$ the result is immediate: $\mathcal{P}(\tau_w)=\frac{1}{\pi \sqrt{(b-\tau_w)(\tau_w-a)}}$. The case $N=2$ is more complicated  but someone can notice
that in this case the Wigner time is  closely related to  the Seebeck coefficient $S_k$ \cite{AdelThesis} so that by using the integrals and techniques exploited in \cite{1stArticle}\cite{AdelThesis} we find the distribution at arbitrary energy ($M=N=2$): 
\begin{eqnarray}
  \mathcal{P}_E(\tau_w)\propto\frac{\mathstrut_2 F_1(\frac{1}{2},\frac{1+\beta}{2},\frac{2+\beta}{2},1-(\frac{\tau_w-\langle\tau_w\rangle}{\alpha})^2)}{(1-(\frac{\tau_w-\langle\tau_w\rangle}{\alpha})^2)^{-\beta/2}} \label{WignerTime}
\end{eqnarray}
$\mathstrut_2 F_1$ is the Gauss hypergeometric function\cite{Wolfram}.
\begin{figure}
 \begin{center}
  \includegraphics{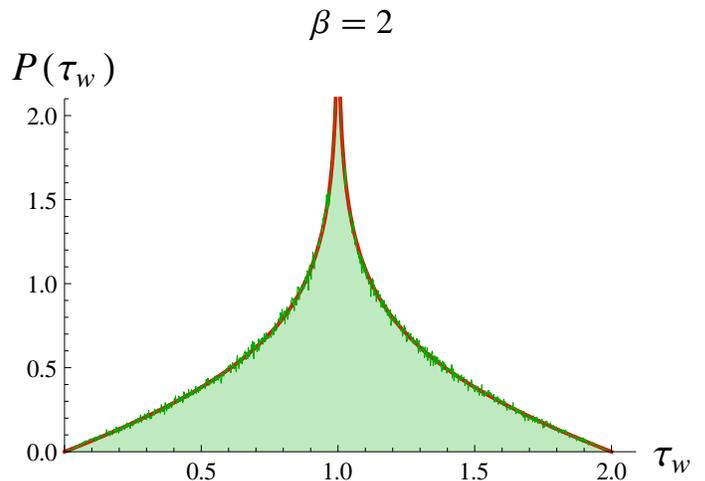}
\caption{Distribution of the Wigner time in the absence of time-reversal symmetry ($\beta=2$). Green plot is obtained numerically whereas the red curve is obtained with Eq.~(\ref{WignerTime}). The 
Fermi energy (here $E_F=0$) lies in the edge of the Hamiltonian spectrum ($\rho(E_F)=2. 10^{-4}$). We used the new self-energy $\Sigma^\prime$ to reach this density at $E_F=0$. The size of the cavity is $N=26$ with 
two conducting modes ($M=2$).  }
\label{NewWignerTime} 
\end{center}
\end{figure}
This distribution is singular and has a logarithmic divergence near the mean Wigner time : $\mathcal{P}_E(\tau_w)\sim C_\beta \log(\frac{|\tau_w-\langle\tau_w\rangle|}{\alpha})$, where $C_\beta$ is a $\beta$-dependent constant.
As for all the observables depending on $\mathcal{S}$ and $\frac{\partial \mathcal{S}}{\partial E}$, at the spectrum edge when $\Gamma \rightarrow 0$, Eq.(\ref{WignerTime}) describes very well the Wigner time distribution even for $N>M=2$ (but $N$ finite).\\
The form of the self energy of a uniform lead implies that the edge of the spectrum is obtained for a Fermi energy lying in the vicinity of the bottom of the conduction band. In order,
to check that the physics discussed here is due to the spectrum edge (and not to the bottom of the band) we suppose that in addition to the uniform semi-infinite lead
there are  stumps\cite{Varbanov} attached to the cavity. This implies a new self energy $\Sigma^\prime=q+\Sigma$. The additional part $q$, due to the stump, can be chosen 
energy-independent in some limits of the coupling term. When $q\gg \Gamma/2$, the level density at $E_F$ becomes $\rho (E_F)\sim \frac{\Gamma}{2 \pi q^2}\ll 1$. This very low density implies 
that at any Fermi energy, we are in the tail of the spectrum. Since  the energy derivative of $\Sigma^\prime$ does  not change from the previous case, all the conclusions
claimed so far remain unchanged and moreover they become true at any Fermi energy: The minimal chaotic cavity gives the distribution of the transport
 coefficients (depending on $\mathcal{S}$ or $\frac{\partial\mathcal{S}}{\partial E}$ of any cavity with a finite $N$  at the spectrum edge. This is successfully tested with
numerical simulations: Fig. \ref{NewWignerTime} shows the comparison between simulation and the analytical result of the distribution of the Wigner time. The simulation is based on sampling Hamiltonians giving rise 
to scattering matrices from the Circular unitary ensemble ($\beta=2$) according to the new self energy $\Sigma^\prime$. The parameter $q$ is chosen quite large to reach the edge of the Hamiltonian spectrum at $E_F=0$.
 The distribution is symmetric around the the Wigner mean time $\bar{\tau}_w=\frac{2}{\Gamma}$.
 
As a conclusion, we proposed a model of minimal chaotic cavities with scattering matrices from circular ensembles and gave  the exact distribution of the proper delay times and the Wigner time.   We showed that 
this model is pertinent to describe in general for $N>M$, the statistics of all the physical quantities depending on $\mathcal{S}$ or/and $\frac{\partial \mathcal{S}}{\partial E}$ at the edge of the Hamiltonian spectrum.

\begin{acknowledgements}
The author is grateful to J. L. Pichard and K. Muttalib for introducing him to this subject and to RMT in general. He would like to thank G. Fleury for valuable discussions and remarks.
M. Albert is kindly acknowledged for his pertinent remarks. \\
The author acknowledges partial support of the R\'egion Basse Normandie.
\end{acknowledgements}

\appendix
\section{Density of proper delay times} \label{AppenA}
The density of proper delay times is defined as follows: 
\begin{eqnarray}
 \rho(\tau)=\sum_i \langle \delta(\tau-\tau_i)\rangle
\end{eqnarray}
this means:
\begin{eqnarray}
 \rho(\tau)&=&\sum_i \int \delta(\tau-\tau_i) \mathcal{P}(\{\tau_k\}) \prod_{k=1}^M d\tau_k  
\end{eqnarray}
Using the parametrization leading to Eq.(\ref{Eigenenergies}) and the equivalence between the proper delay times we can write: 
\begin{eqnarray}
 \rho(\tau) = M\int \delta(\tau-f(\epsilon_i)) \mathcal{P}(\{\epsilon_k\}) \prod_{k=1}^M  d\epsilon_k
\end{eqnarray}
$f(x)=\frac{x^2}{1+x^2}$,   ($\tau_i=\frac{\epsilon_i^2}{1+\epsilon_i^2}$). $f^{-1}(x)=\sqrt{\frac{x}{1-x}}.$\\
The properties of the Dirac delta function allow to write:
\begin{align}
\rho(\tau)  &=& 2M \int \frac{\delta(\epsilon_i-f^{-1}( \tau))}{|f^\prime(f^{-1}(\tau))|} \mathcal{P}(\{\epsilon_k\}) \prod_{k=1}^M  d\epsilon_k \nonumber\\
            &=&\frac{1}{(1-\tau)^{3/2} \tau^{1/2}} \underbrace{M\int \delta(\epsilon_i-f^{-1}( \tau)) \mathcal{P}(\{\epsilon_k\}) \prod_{k=1}^M  d\epsilon_k}_{\rho_{\mathcal{L}E}(f^{-1}(\tau))} \nonumber\\
            &=& \frac{ \rho_{\mathcal{L}E}(f^{-1}(\tau))}{(1-\tau)^{3/2} \tau^{1/2}} \label{EquationA4}
\end{align}
where $\rho_{\mathcal{L}E}$ is the density of eigenvalues of Lorentzian ensembles which is given by (in the half filling limit)\cite{BrouwerLorentz}:
\begin{eqnarray}
\rho_{\mathcal{L}E}(\epsilon)=\frac{M}{\pi} \frac{1}{1+\epsilon^2} \label{EquationA5}
\end{eqnarray}
From Eqs.(\ref{EquationA4}) and (\ref{EquationA5}), we obtain (at $E_F=0$):
\begin{eqnarray}
 \rho(\tau)=\frac{M}{\pi} \frac{1}{\sqrt{\tau(1-\tau)}}
\end{eqnarray}
We notice that the density of eigentimes do not depend on the symmetry class which goes back to the fact that the Lorentzian density of eigenvalues is $\beta$-independent.
The generalization of this result to any arbitrary energy can be obtained following the same steps, or just by using Eq.(\ref{InvarianceFormula}). We finally get:
\begin{eqnarray}
\rho(\tau)=\frac{M}{\pi} \frac{1}{\sqrt{(b-\tau)(\tau-a))}}
\end{eqnarray}

\section{Wigner mean time}\label{AppenB}
We show here how we calculate the Wigner mean time. We will use the fact that the marginal distribution of the eigenphases (phases of  $\mathcal{S}$ eigenvalues) is uniform.
We have:
\begin{align}
 \langle \tau_w\rangle&=& \int  (\tau_1+\tau_2\dots +\tau_M) \mathcal{P}(\{\tau_i\}) \prod_{i=1}^M d\tau_i\\
                      &=& M\int  \tau_1 \mathcal{P}(\{\tau_i\}) \prod_{i=1}^M d\tau_i \\
                      &=& M\int  \tau_1(\varphi_1) \mathcal{P}(\{\varphi_i\}) \prod_{i=1}^M d\varphi_i \\
                      &=& M\int_{-\pi}^{\pi} \tau_1(\varphi_1)  \frac{d\varphi_1}{2\pi} \label{EquationB4}
\end{align}
because $\int \mathcal{P}(\{\varphi_i\}) \prod_{i \ne 1}d\varphi_i =\frac{1}{2\pi} $\cite{Dyson}. We recall that for circular ensembles, we have $\mathcal{P}(\{\varphi_i\})\propto \prod_{i<j}|e^{i\varphi_i}-e^{i\varphi_j}|^\beta$. \\
From Eq.(\ref{GeneralQ}) we deduce the relation between $\tau_i$ and $\varphi_i$: 
\begin{align}
 \tau_i(\varphi_i)&=&-\frac{\alpha}{2} \frac{e^{+i(\varphi_i+\theta)}+e^{-i(\varphi_i+\theta)}}{2}+\frac{1}{\Gamma} \nonumber\\   
               &=& -\frac{\alpha}{2} \cos(\varphi_i+\theta)+\frac{1}{\Gamma} \label{EquationB5}
\end{align}
Eqs. (\ref{EquationB5}) and (\ref{EquationB4}) lead to the final result:
\begin{eqnarray}
 \bar{\tau}_w\equiv \langle \tau_w\rangle&=& \frac{M}{\Gamma}.
\end{eqnarray}

\section{The second cumulant}\label{AppenC}
We would like to calculate the second cumulant of the Wigner time defined as :
\begin{eqnarray}
\langle \delta\tau_w^2 \rangle = \langle (\tau_w-\bar{\tau}_w)^2 \rangle
\end{eqnarray}
It is better to do the calculation using the eigenphases representation as in Eq.(\ref{EquationB5}) :

\begin{eqnarray}
\langle \delta\tau_w^2 \rangle = \frac{\alpha^2}{4}\langle(\sum_i\cos(\varphi_i+\theta))^2 \rangle
\end{eqnarray} 
First, we notice that a translation $\varphi_i+\theta\rightarrow \varphi_i$ does not change the result.
Second, the expansion of this expression with the use of the symmetries between eigenphases allow to write ($i$, $j$ are arbitrary with $i\ne j$):
\begin{eqnarray}
\langle \delta\tau_w^2 \rangle = \frac{\alpha^2}{4}  (N\langle(\cos^2(\varphi_i) \rangle+N(N-1)\langle(\cos(\varphi_i) \cos(\varphi_j) \rangle \nonumber \\
\end{eqnarray} 
 The first term is simple since the marginal distribution of one eigenphase is uniform\cite{Dyson}:
\begin{eqnarray}
 R_1(\varphi)=\int \delta(\varphi-\varphi_i) \mathcal{P}(\{\varphi_k\}) \prod_{k=1}^M d\varphi_k=\frac{1}{2\pi }
\end{eqnarray}
this gives :
\begin{eqnarray}
 \langle(\cos^2(\varphi_i) \rangle=\frac{1}{2}.
\end{eqnarray}
The second term involves the density-density correlation function. We have: 
\begin{eqnarray}
 R_2(\varphi,\varphi^\prime)=\int \delta(\varphi-\varphi_i) \delta(\varphi^\prime-\varphi_j) \mathcal{P}(\{\varphi_k\}) \prod_{k=1}^M d\varphi_k \nonumber \\
\end{eqnarray}
with ($i\ne j$).

\begin{eqnarray}  
   \begin{array}{r c l}
      R_2(\varphi,\varphi^\prime)  & = & \langle\delta(\varphi-\varphi_i) \delta(\varphi^\prime-\varphi_j) \rangle  \\
         & = & \frac{1}{N(N-1)}\sum_{i\ne j}\langle\delta(\varphi-\varphi_i) \delta(\varphi^\prime-\varphi_j) \rangle \\
       & = & \frac{1}{N(N-1)} (N^2 \overline{\rho(\varphi) \rho(\varphi^\prime)}-N\delta(\varphi-\varphi^\prime) \overline{\rho(\varphi)}
   \end{array}
\end{eqnarray}
and $\rho(\varphi)=\frac{1}{N} \sum_i \delta(\varphi-\varphi_i)$. We defined $\overline{A}\equiv\langle A\rangle$.\\
One can notice that $\overline{\rho(\varphi)}=R_1(\varphi)$.
We can express $\rho(\varphi)$ otherwise
\begin{eqnarray}  
   \begin{array}{r c l}
      \rho(\varphi)  & = & \frac{1}{N} \sum_{i=1}^N \delta(\varphi-\varphi_i)  \\
         & = & \frac{1}{2\pi N} \sum_i^N \sum_{n=-\infty}^{n=+\infty} e^{i n (\varphi-\varphi_i)} \\
       & = & \frac{1}{2\pi N} \sum_{n=-\infty}^{n=+\infty} t_n e^{i n \varphi} 
   \end{array}
\end{eqnarray}
where we put  $t_n=Tr(S^{\dagger n})=\sum_i e^{-i n \varphi_i}$ . The density-density correlation function is :
\begin{eqnarray}
  \overline{\rho(\varphi) \rho(\varphi^\prime)}=\left(   \frac{1}{2\pi N} \right)^2 \sum_n \overline{|t_n|^2} e^{i n (\varphi-\varphi^\prime)}
\end{eqnarray}
We want to express the following integral:
\begin{eqnarray}
 \langle \cos(\varphi) \cos(\varphi^\prime) \rangle=\int \cos(\varphi) \cos(\varphi^\prime)R_2(\varphi,\varphi^\prime) d\varphi d\varphi^\prime \nonumber \\
\end{eqnarray}
This will be done term by term:
\begin{align}
 I_1&=&\int \cos(\varphi) \cos(\varphi^\prime) \overline{\rho(\varphi) \rho(\varphi^\prime)} d\varphi d\varphi^\prime \nonumber \\
                                       &=&\left(\frac{1}{2\pi N}\right)^2\sum_n \overline{|t_n|^2} \int \cos(\varphi) \cos(\varphi^\prime) e^{i n (\varphi-\varphi^\prime)}\nonumber\\ 
                                       &=&\left(\frac{1}{2\pi N}\right)^2\sum_n \overline{|t_n|^2} |\int \cos(\varphi) e^{i n \varphi} d\varphi |^2\nonumber\\
                                       &=&\left(\frac{1}{2\pi N}\right)^2\sum_n \overline{|t_n|^2} |\int \frac{(e^{i (n+1) \varphi}+e^{i (n-1) \varphi})}{2} d\varphi |^2 \nonumber\\
                                       &=&\left(\frac{1}{2\pi N}\right)^2 \sum_n \overline{|t_n|^2}  (2\pi )^2 |\frac{\delta_{n+1,0}+\delta_{n-1,0}}{2} |^2\nonumber\\
                                       &=&\frac{1}{ 4 N^2} (\overline{|t_1|^2}+\overline{|t_{-1}|^2} \nonumber\\
                                       &=& \frac{\overline{|t_1|^2}}{ 2 N^2}\nonumber\\ 
\end{align}
The second term to evaluate leads to :
\begin{align}
 I_2&=&\int \cos(\varphi) \cos(\varphi^\prime) \delta(\varphi-\varphi^\prime) \overline{\rho(\varphi)} d\varphi d\varphi^\prime \nonumber\\
                                       &=& \int \cos^2(\varphi) \overline{\rho(\varphi)} d\varphi \nonumber\\
                                       &=& \frac{1}{2}\nonumber\\
\end{align}
All in all, we obtain the following result correct for the three ensembles($\beta=1,\beta=2$ and $\beta=4 $) :

\begin{eqnarray}
\langle \delta\tau_w^2 \rangle =\frac{\alpha^2}{8} \overline{|t_1^2|}, 
\end{eqnarray}
for $\beta=1$ $\overline{|t_1|^2}=1$ and for $\beta=2$ the form factor is $\overline{|t_1|^2}=\frac{2 N}{N+1}$. In the case $\beta=4$, the 
form factor  $\overline{|t_1|^2} =\frac{N}{2N-1}$.






\begin{thebibliography}{25}
\bibitem{Remark} In all the paper, we take  $t_0=1$.
\bibitem{Wolfram} $\mathstrut_2 F_1$ is used as defined and implemented in Mathematica Wolfram
\end{thebibliography}
\end{document}